\documentclass[prd,superscriptaddress,amsfonts,amssymb,amsmath,showpacs,onecolumn]{revtex4-2}
\usepackage{bm}
\usepackage{amsfonts}
\usepackage{latexsym}
\usepackage[utf8]{inputenc}
\usepackage[T1]{fontenc}
\usepackage{graphicx}
\usepackage{amsmath}
\usepackage{mathrsfs}  
\usepackage{palatino}
\usepackage{mathpazo}
\usepackage{textcomp}
\usepackage{float}
\usepackage{booktabs}
\usepackage{dcolumn}
\usepackage{multirow}
\usepackage{hyperref}
\hypersetup{colorlinks,citecolor=blue}
\usepackage{amsmath}
\usepackage{xcolor}
\usepackage{orcidlink}
\usepackage[caption=false]{subfig}
\usepackage{commath}
\def\jnl@style{\it}
\def\aaref@jnl#1{{\jnl@style#1}}

\def\aaref@jnl#1{{\jnl@style#1}}

\def\aj{\aaref@jnl{AJ}}                   
\def\apj{\aaref@jnl{ApJ}}                 
\def\apjl{\aaref@jnl{ApJ}}                
\def\apjs{\aaref@jnl{ApJS}}               
\def\apss{\aaref@jnl{Ap\&SS}}             
\def\aap{\aaref@jnl{A\&A}}                
\def\aapr{\aaref@jnl{A\&A~Rev.}}          
\def\aaps{\aaref@jnl{A\&AS}}              
\def\mnras{\aaref@jnl{Mon.~Not.~Roy.~Astron.~Soc.}}             
\def\prd{\aaref@jnl{Phys.~Rev.~D}}        
\def\prc{\aaref@jnl{Phys.~Rev.~C}}  
\def\prl{\aaref@jnl{Phys.~Rev.~Lett.}}    
\def\qjras{\aaref@jnl{QJRAS}}             
\def\skytel{\aaref@jnl{S\&T}}             
\def\ssr{\aaref@jnl{Space~Sci.~Rev.}}     
\def\zap{\aaref@jnl{ZAp}}                 
\def\nat{\aaref@jnl{Nature}}              
\def\aplett{\aaref@jnl{Astrophys.~Lett.}} 
\def\apspr{\aaref@jnl{Astrophys.~Space~Phys.~Res.}} 
\def\physrep{\aaref@jnl{Phys.~Rep.}}      
\def\physscr{\aaref@jnl{Phys.~Scr}}       
\def\commat{\aaref@jnl{Comm.~Math.~Phys.}}              
\def\science{\aaref@jnl{Science}}               
\def\cqg{\aaref@jnl{Classical Quant.~Grav.}}            
\def\jpcs{\aaref@jnl{JPCS}}                                     
\def\ijmpd{\aaref@jnl{Int.~J.~Mod.~Phys.~D}}                    
\def\grg{\aaref@jnl{Gen.~Relat.~Gravit.}}               
\def\rpp{\aaref@jnl{Rep.~Prog.~Phys.}}          
\def\npa{\aaref@jnl{Nucl.~Phys.~A}}        
\def\lrr{\aaref@jnl{Living Rev.~Rel.}}                   
\def\jcap{\aaref@jnl{J.~Cosmology Astropart.~Phys.}}    
\def\rmp{\aaref@jnl{Rev.~Mod.~Phys.}}   
\def\epjc{\aaref@jnl{Eur.~Phys.~J.~C}}

\allowdisplaybreaks[1]

\begin{document}

\color{black}       

\title{Non-singular T-K axion stars with/without the
dynamical bosonic field in the presence of negative $\Lambda$ term}

\author{Oleksii Sokoliuk\orcidlink{0000-0003-4503-7272}}
\email{oleksii.sokoliuk@mao.kiev.ua}
\affiliation{Main Astronomical Observatory of the NAS of Ukraine (MAO NASU),\\
Kyiv, 03143, Ukraine}
\affiliation{Astronomical Observatory, Taras Shevchenko National University of Kyiv, \\
3 Observatorna
St., 04053 Kyiv, Ukraine}

\author{Alexander Baransky\orcidlink{0000-0002-9808-1990}}
\email{abransky@ukr.net}
\affiliation{Astronomical Observatory, Taras Shevchenko National University of Kyiv, \\
3 Observatorna
St., 04053 Kyiv, Ukraine}
\author{P.K. Sahoo\orcidlink{0000-0003-2130-8832}}
\email{pksahoo@hyderabad.bits-pilani.ac.in}
\affiliation{Department of Mathematics, Birla Institute of Technology and
Science-Pilani,\\ Hyderabad Campus, Hyderabad-500078, India.}

\date{\today}
\begin{abstract}

In the present work authors derive exact solutions for the relativistic compact stars in the presence of two fields axion (Dante’s Inferno model) and with/without
the complex scalar field (with the quartic self-interaction) coupled to gravity. The matter source is assumed as the perfect fluid one, and we also use barotropic/MIT bag EoS
to derive energy density and anisotropic/isotropic pressures from Einstein Field Equations (EFE’s). For simplicity, as the metric potentials we use Tolman–Kuchowicz metric
potentials, which are non-singular and physically acceptable. Unknown variables for
the T-K metric potentials were derived from the junction conditions. To examine the
redshift function, adiabatic index and energy, causality conditions we used such compact stars as PSR J1416-223, PSR J1903+32, 4U 1820-30, Cen X-3, EXO 1785-248,
SAX J1808.4365.

\end{abstract}

\maketitle
\section{Introduction} \label{sec:1}
The study of relativistic stellar objects in the General Theory of Relativity (further - GR) and in the modified theories of gravity is considered as one of the most popular and promising during the last decades. There was written a large amount of papers on the subject of relativistic compact stars, such as \cite{PhysRevD.97.044022,PhysRevD.95.104019,Ovalle2018a,Ovalle2018b,Ovalle2015} (and references therein).
Compact stars are the massive and very small in radius stellar objects that is the final stage of the stellar evolution. Compact stars are separated on the white dwarfs, neutron stars and more exotic objects like strange stars (made from the strange flavored quark matter). For example, neutron stars are compact objects which are boosted by their neutron degeneracy pressure against the attraction of gravity. On other hand, white
dwarfs are usually boosted by the electron degeneracy pressure against the gravity.
Nowadays, it is of special interest to study such kind of objects, because compact stars are laboratories, in which we could investigate the strongly coupled gravitational fields and nuclear processes. 
Finally, it is also worth to notice that as we said previously, during the past few years relativistic stellar objects were also investigated in many different (viable an not) modified gravity theories. 
For example, in the framework of $f(\mathcal{R})/f(\mathcal{R},T)$ gravities following works were published:
\cite{Sharif:2021emv,Das2016CompactSI,Biswas:2020gzd,Kumar:2021vqa,Bhar:2021iog,Rej:2021ngp}. In turn, for the teleparallel gravity (and their modifications) there is also exist a couple of works, namely \cite{Maurya:2021mqx,Sharif:2021mwf,Lin:2021ijx,Zubair:2021qvw,Solanki:2021fzo,deAraujo:2021pni} (and references therein).

In our paper, we will study the compact stars with two coupled to gravity axion pseudoscalar fields and (optionally) with the one dynamical bosonic field. Metric potentials of the compact star line element are considered to be of the Tolman–Kuchowicz kind (this kind of metric potentials are well studied and physically acceptable, non-singular).
\subsection{Dante's Inferno model}
The model that we consider with the two free axion pseudoscalar fields coupled to gravity is called Dante's inferno model (one of the axion monodromy models) and was firstly presented by \cite{PhysRevD.81.103535}. This model potentially could describe inflation and could be embeded into the string theory as a generalization of axion monodromy model. In this model high-scale large-field inflationary dynamics takes place within a region of field
space which is parametrically subplanckian in diameter \cite{PhysRevD.81.103535}. Dante's inferno could give a fresh view of the Lyth bound (which provides the upper limit of the gravitational waves that could be produced during the period of inflation), and thus is very interesting case to investigate for the relativistic compact objects.
\subsection{Why Anti-de Sitter?}
\begin{figure}[!h]
\centering
\includegraphics[width=0.4\textwidth]{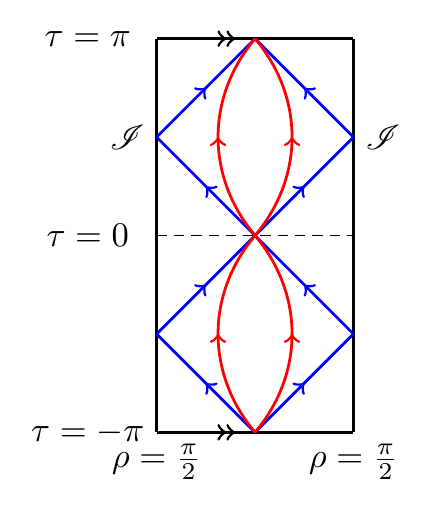}
\qquad
\includegraphics[width=0.4\textwidth]{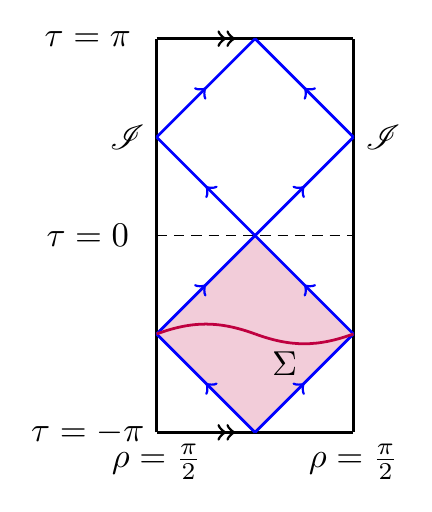}
\caption{Conformal structure of the Anti-de Sitter spacetime (figure taken from the \cite{Ambrus2018})}
\label{fig:111}
\end{figure}

Anti-de Sitter spacetime is one of the simplest and most symmetric solutions of Einstein Field Equations including the negative cosmological constant and it is widely used in the GR, modified field theories and string theory, braneworld models like Randall-Sundrum one. In the string theory and high energy physics AdS spacetime is the very hot topic among the researchers due to the Maldacena's AdS/CFT (Conformal Field Theory) correspondence, suggesting that fundamental particle interactions may be described in geometrical terms \cite{Sokolowski2016}. Also, AdS spacetime is widely used as the AdS$_5$ bulk with the embeded 3-branes in it (if we consider RS models). Also, Anti-de Sitter spacetime could be a viable background, that describes current expansion rate, BAO/CMB data with high precision \cite{Sen:2021wld}. So it will be helpful to probe the compact stars in the presence of negative cosmological constant. Additionally, on the Figure (\ref{fig:111}) we illustrated the conformal structure of the Anti-de Sitter spacetime. On this graph $\mathscr{I}$ means null-like infinity, $\tau$ is the cyclic AdS time, red lines are time-like geodesics, blue lines are null-like geodesics, $\Sigma$ is the Cauchy curve, shaded region is the so called Wheeler-De Witt (WDW) patch.
\subsection{Article organisation}
Our article is organised as follows: in the Section (\ref{sec:1}) we provide an introduction into the topic of relativistic stellar objects, Dante's Inferno model and why we consider Anti-de Sitter. In the Section (\ref{sec:2}) we derive Einstein Field Equations for the axion stars, probe their energy, causality conditions and stability through adiabatic perturbations and TOV equation with the barotropic and MIT bag Equation of State (EoS). In the Section (\ref{sec:3}) we introduce the dynamical complex scalar field and do the same procedure as was done in the previous section. Finally, we provide the concluding remarks about the key topics of our study in the last Section (\ref{sec:4}).
\section{Einstein Field equations for axion star} \label{sec:2}
To derive the suitable Einstein Field equations (EFE's), we firstly must specify the so-called Einstein-Hilbert (EH) action integral. If we assume the GR classical gravity and as a matter source two axion pseudoscalar fields, then the EH action given as follows \cite{PhysRevD.97.124052}:
\begin{equation}
    \mathcal{S}[g,\Gamma,\psi_I,\phi,\Psi_i] = \int_\mathcal{M}d^Dx \sqrt{-g}\bigg[\kappa(\mathcal{R}-2\Lambda)-\frac{1}{2}\sum^2_{I=1}(\partial^\mu\psi_I\partial_\mu\psi_I))+\mathcal{L}(\Psi_i)\bigg]
\end{equation}
where $g=\det g_{\mu\nu}=\prod^{D-1}_{\mu,\nu=0}g_{\mu\nu}$ is the metric determinant, $\{\psi_1,\psi_2\}$ are two axion scalar fields and finally $\mathcal{L}(\Psi_i)$ is the Lagrangian density of the perfect fluid matter fields $\Psi_i$. Then, the set of EFE's is (in the limit $D\to4$):
\begin{equation}
    \kappa(G_{\mu\nu}+\Lambda g_{\mu\nu})=\frac{1}{2}(T^\psi_{\mu\nu}+T_{\mu\nu}^{\mathrm{M}})
\end{equation}
\begin{equation}
    \Box \psi_I=0, \quad  \forall  I
    \label{eq:4}
\end{equation}
where $g^{\mu\nu}\nabla_\mu\nabla_\nu$ is the d'Alambert operator. Generally, stress-energy-momentum tensor could be written as a variation of the matter fields Lagrangian density:
\begin{equation}
    T_{\mu\nu} = -\frac{2}{\sqrt{-g}}\frac{\delta(\sqrt{-g}\mathcal{L}(\Psi_i))}{\delta g^{\mu\nu}}
\end{equation}
Also, stress-energy tensor for scalar fields is \cite{PhysRevD.97.124052}:
\begin{equation}
    T^\psi_{\mu\nu}=\sum^2_{I=1}(\partial_\mu\psi_I\partial_\nu\psi_I-\frac{1}{2}g_{\mu\nu}(\partial^\mu\psi_I\partial_\mu\psi_I))
\end{equation}
Consequently, we assume the $D$ dimensional spherically symmetric spacetime line element of form:
\begin{equation}
    ds^2_D = e^{\Phi(r)}-e^{\zeta(r)}dr^2-r^2d\Omega^2_{D-2}
\end{equation}
where the $D-2$ dimensional unit sphere line element:
\begin{equation}
    \begin{gathered}
    d\Omega^2_{D-2}=d\theta_1^2+\sin^2\theta_1d\theta_2^2+\sin^2\theta_1\sin^2\theta_2d\theta_3^2+...+\bigg(\prod^{D-3}_{j=1}\sin^2\theta_j\bigg)d\theta^2_{D-2}
    \end{gathered}
\end{equation}
Stress energy tensor for the prefect fluid matter source in our case reads
\begin{equation}
    T_{\mu}^{\nu} =  (\rho+p_t)U_\mu U^\nu - p_t\delta^\nu_\mu + (p_r-p_t)V_\mu V^\nu
\end{equation}
Here, $\rho$ is matter energy density, $p_r$, $p_t$ is radial and tangential pressures respectively, $V_\mu$ is unitary radial spacelike vector, $U_\mu$ is four velocity. Further we will restrict our analysis to the $(3+1)$ dimensions. 
\subsection{Axion fields}
By assuming that the axion fields $\psi_I$ depend only on the coordinates $\theta$ and $\phi$ and by solving the field equation (\ref{eq:4}) we get the following solutions:
\begin{equation}
    \psi_1  = c_1 + c_2 \theta
\end{equation}
\begin{equation}
    \psi_2  = c_1 + c_2 \phi
\end{equation}
where $c_1$ and $c_2$ are variables of integration. In the further investigation, we will assume that $c_1\to0$ and $c_2\to\lambda$.

\subsection{Tolman–Kuchowicz metric}
We have chosen the Tolman–Kuchowicz (T-K) spacetime because of the fact that Tolman–Kuchowicz metric potentials are well studied and non-singular. Kuchowicz potential is given by \cite{Kuchowicz1968}:
\begin{equation}
    e^{\zeta(r)} = e^{Cr^2+2\ln D} 
\end{equation}
and Tolman-like potential \cite{Tolman1939}:
\begin{equation}
  e^{\Phi(r)} = e^{\ln(1+Ar^2+Br^4)}
\end{equation}
where $A$, $B$, $C$ and $D$ are arbitrary constants. Regularity of the metric potentials at the origin $r=0$ could be easily checked. Firstly, at $r=0$, $e^{\Phi(0)}=e^{2\ln D}$ and $e^{\zeta(0)}=0$. Also,
\begin{equation}
    (e^{\Phi(r)})' = (2Ar+4Br^3)\rvert_{r=0}=0
\end{equation}
\begin{equation}
    (e^{\zeta(r)})'=(2CD^2e^{Cr^2}r)\rvert_{r=0}=0
\end{equation}
So both metric potentials are regular at the origin and there is no singularities present.
\subsection{EFE's in the T-K spacetime}
Thus, the (modified) Einstein field equations are:
\begin{equation}
\begin{gathered}
\rho=-\frac{1}{r^2}\bigg[2 \bigg(\frac{e^{-C r^2} \left(2 C r^2-1\right)}{D^2}-\Lambda  r^2+1\bigg)\bigg]
\end{gathered}
\end{equation}
\begin{equation}
\begin{gathered}    
    p_r =- \frac{1}{r^2}\bigg[2 \bigg(\frac{e^{-C r^2} \left(3 A r^2+5 B r^4+1\right)}{D^2 \left(A r^2+B r^4+1\right)}+\Lambda  r^2-1\bigg)\bigg]
   \end{gathered}
\end{equation}
\begin{equation}
    \begin{gathered}
    p_t =-\bigg[2 e^{-C r^2} \left(-2 r^4 \left(A^2 C-3 A B+2 B C\right)+r^2 \left(A^2-3 A C+8 B\right)+B r^6 (4 B-5 A C)+2 A-3 B^2 C r^8-C\right)\bigg]\\\bigg/\bigg[D^2 \left(A r^2+B r^4+1\right)^2\bigg]+2 \Lambda +\frac{\lambda^2 }{2 r^2}
    \end{gathered}
\end{equation}
\begin{equation}
    \begin{gathered}
    p_t =-\bigg[e^{-C r^2} \left(-4 r^4 \left(A^2 C-3 A B+2 B C\right)+2 r^2 \left(A^2-3 A C+8 B\right)+2 B r^6 (4 B-5 A C)+4 A-6 B^2 C r^8-2 C\right)\bigg]\\\bigg/\bigg[D^2 \left(A r^2+B r^4+1\right)^2\bigg]+2 \Lambda +\frac{ \lambda^2  \csc
   ^2(\theta)}{2r^2}
    \end{gathered}
\end{equation}
Also,  if we assume that our axion star live in the AdS spacetime, then we could apply proper transformation  $\Lambda\to-\frac{(D-1)(D-2)}{2l^2}\to-3/l^2<0$ where $l$ is the AdS radius. For the sake of simplicity, we will investigate only the equatorial region of the axion star, and thus because of the spherical symmetry $\theta=\pi/2$.
\subsection{Junction conditions}
In order to evaluate the constants ($A$, $B$, $C$, $D$), we could apply the junction conditions, which provides smooth junction between the exterior and interior spacetimes at the hypersurface $(\Sigma:r=R)$. f{In our paper, the external vacuum spacetime is completely described by the Schwarzschild-AdS metric}:
\begin{equation}
    ds^2 = f(R)dt^2 - f(R)^{-1}dr^2 - R^2 d\theta^2 - R^2 \sin^2 \theta d\phi^2
\end{equation}
where the function $f(R)$ is defined below
\begin{equation}
    f(R)=\frac{1-2M}{R}-\frac{\Lambda R^2}{3}
\end{equation}
The junction hypersurface is described by the following spacetime:
\begin{equation}
    ds^2 = dT^2 - R^2(d\theta^2+\sin^2\theta d \phi^2)
\end{equation}
where $T$ is the proper time on the $\Sigma$. Then, the curvature tensor is:
\begin{equation}
    K_{ij}^{\pm} = -\frac{\partial^2z^\gamma_{\pm}}{\partial \eta^i \eta^j}n^\pm_\gamma - \Gamma^\gamma_{\delta\mu}\frac{\partial z_\pm^\delta}{\partial \eta^i}\frac{\partial z_\pm^\mu}{\partial \eta^j}n^\pm_\gamma
\end{equation}
Here $-$ denotes the interior spacetime and $+$ exterior, $z_\pm^\mu$ are the coordinates of the both interior/exterior regions, $\Gamma^\gamma_{\;\delta\mu}$ is the so-called Levi-Civita affine connection, $n^\pm_\mu$ - normal to the boundary, $\eta^i$ - coordinates on the hypersurface $\Sigma$. Then, plugging this all up and following \cite{Majid2020}, we could obtain the metric potentials on the hypersurface:
\begin{equation}
    e^{\Phi(R)} = \frac{1-2M}{R}-\frac{\Lambda R^2}{3}, \quad e^{-\lambda(R)} =\frac{1-2M}{R}-\frac{\Lambda R^2}{3}, \quad p\rvert_\Sigma=0
\end{equation}
where $R$ is the compact star radius and $M$ is its mass. Because of the necessary condition on the hypersurface $\Sigma$ for radial pressure to vanish and from the hypersurface metric potentials, we have:
\begin{equation}
\begin{gathered}
A = -\frac{B R^5+2 M+R-1}{R^3}-\frac{\Lambda }{3}
\end{gathered}
\label{eq:22}
\end{equation}
\begin{equation}
    C= \frac{1}{R^2}\log \biggl(-\frac{3 R}{D \bigl(6 M+\Lambda  R^3-3\bigr)}\biggr)
    \label{eq:25}
\end{equation}
\begin{table}[]
\centering
\resizebox{0.8\textwidth}{!}{%
\begin{tabular}{llllllll}
\hline
Star            & $M (M_\odot)$ & $R$(km) & $A$         & $B$            & $C$          & $D$        & Reference                                                \\ \hline
PSR J1416-2230  & 1.97          & 9.69    & $1.95284$  & $-0.01$ & $-0.0380214$  & $0.379502$ & \cite{2010Natur.467.1081D}              \\
PSR J1903+327   & 1.667         & 9.438   & $1.90476$ & $-0.01$ & $-0.0403817$ & $0.410789$ & \cite{2011MNRAS.412.2763F}              \\
4U 1820-30      & 1.58          & 9.1     & $1.84304$ & $-0.01$ & $-0.0436963$ & $0.451497$ & \cite{2010}                             \\
Cen X-3         & 1.49          & 8.50    & $1.73956$ & $-0.01$ & $-0.050138$ & $0.519736$ & \cite{2011ApJ...730...25R}              \\
EXO 1785-248    & 1.3           & 8.99    & $1.82278$ & $-0.01$ & $-0.0448271$ & $0.464362$ & \cite{Ozel2008}                         \\
SAX J1808.43658 & 1.32          & 4.14    & $1.25285$ & $-0.01$      & $-0.153508$  & $0.829487$ & \cite{10.1111/j.1365-2966.2009.14562.x} \\ \hline
\end{tabular}%
}
\caption{T-K spacetime coefficients $A$, $B$, $C$ and $D$ for some compact stars (we assumed that the AdS radius is $l=1$ and thus $\Lambda=-3$, also we consider the case with the negative $B=-1$)}
\label{tab:1}
\end{table}

Parameter $D$ could be derived from the condition $p_r(R)=0$:
\begin{equation}
\begin{gathered}
D=\frac{\left(6 M+\Lambda  R^3-3\right) \left(2 B R^5+6 M-\Lambda  R^3+4 R-3\right)}{R \left(\Lambda  R^2+1\right) \left(-6 M+\Lambda  R^3-6 R+3\right)}
   \end{gathered}
   \label{eq:255}
\end{equation}
Then, we left up with only one T-K unknown variable.
On the Table (\ref{tab:1}) we placed the numerical solutions of equations (\ref{eq:22}), (\ref{eq:25}) and (\ref{eq:255}) for some compact stellar objects with known mass and radius. As we noticed, for this stars, metric potentials are physically viable and non-singular for negative $B$, if values of $B$ are positive, solutions become complex.
\subsection{Barotropic axion star}
\subsubsection{Energy conditions}
For the model to be physically acceptable, all the energy conditions,
namely Weak Energy Condition (WEC), Null Energy Condition
(NEC), Strong Energy Condition (SEC) and Dominant Energy Condition (DEC) must be satisfied at every point of the axion star spacetime interior. This energy conditions look like:

\begin{itemize}
\item Null Energy Condition (NEC): $\rho +p_{r}\geq 0$ and $\rho +p_{t}\geq 0
$

\item Weak Energy Condition (WEC) $\rho >0$ and $\rho +p_{r}\geq 0$ and $%
\rho +p_{t}\geq 0$

\item Dominant Energy Condition (DEC): $\rho -|p_{r}|\geq 0$ and $\rho
-|p_{t}|\geq 0$

\item Strong Energy Condition (SEC): $\rho +p_{r}+2p_{t}\geq 0$
\end{itemize}

NEC is minimal requirement of WEC and SEC conditions and should be satisfied
always (if NEC is violated, then the so-called exotic matter or in some
cases phantom fluid will appear \cite{Sahoo2019}). To probe this energy conditions, we assume that axion star matter is described by the following barotropic Equation of State:
\begin{equation}
    p_r = \alpha \rho
    \label{eq:26}
\end{equation}
\begin{equation}
    p_t = \beta \rho
    \label{eq:27}
\end{equation}
where $\alpha\land\beta\in(0,1)$. We considered such equation of state, because it could provide a physically valid description of the perfect fluid for any static and spherically symmetric relativistic system \cite{Azreg_A_nou_2015}. To derive non-singular solutions for $\rho$ and $p_r$ we will use $p_t$, because it contains axion field terms. 
\begin{figure}[!htbp]
    \centering
    \includegraphics[width=\textwidth]{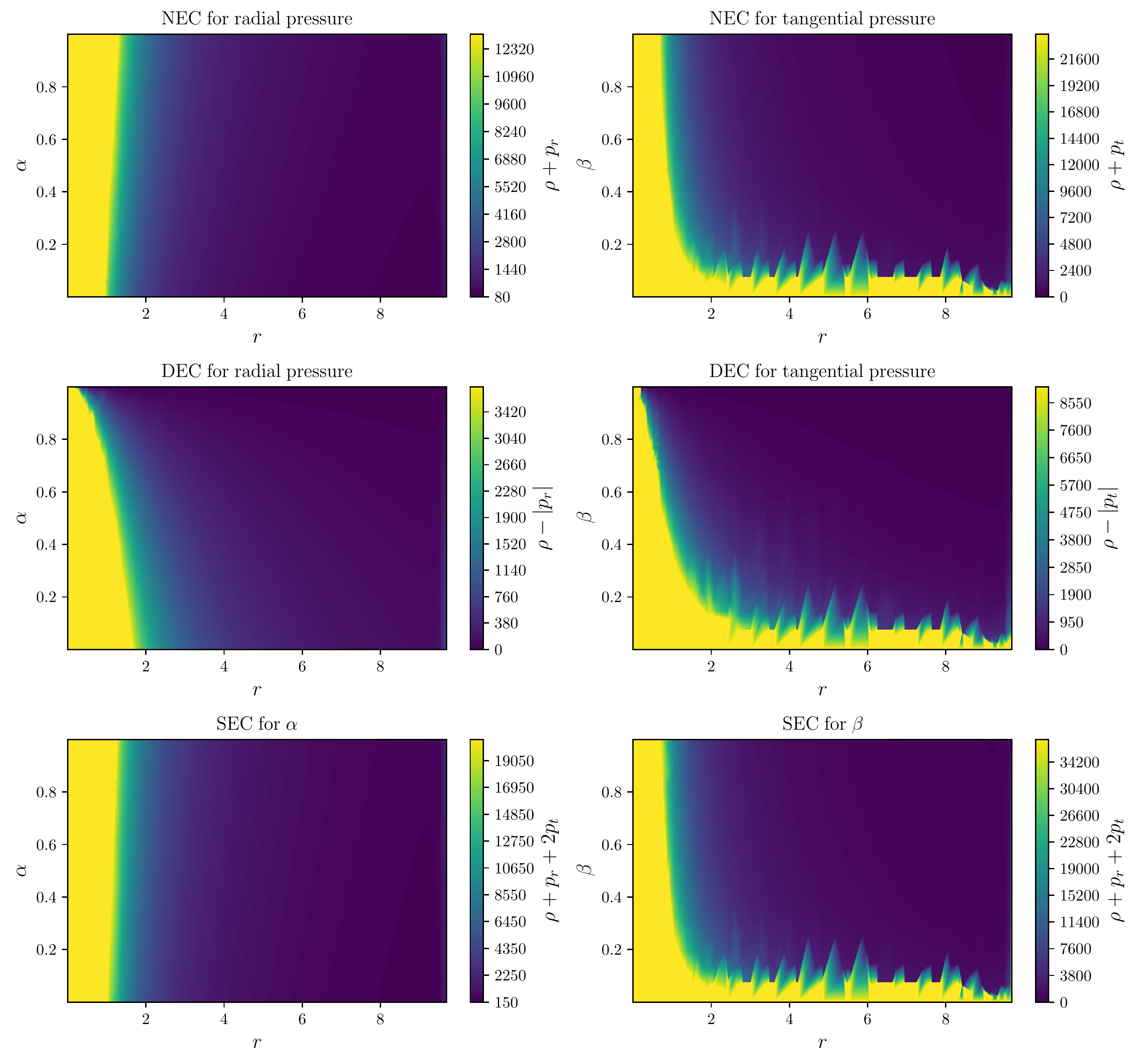}
    \caption{Null, Dominant and Strong Energy Conditions for the PSR J1416-2230 in the presence of two-fields axion}
    \label{fig:1}
\end{figure}
On the Figure (\ref{fig:1}) we represented the NEC, DEC and SEC for PSR J1416-2230 with $M=1.97$ and $R=9.69$. We also assumed that $\lambda=10^2$ and AdS radius is as usual took value $l=1$, $B$ was negative and relatively small. As one could notice, NEC, DEC and SEC was satisfied in the axion star interior for every $\alpha$ in positive bounds $(0,1)$. But, if $|\lambda|$ is not relatively big ($|\lambda|\ll0$), Null Energy condition will be violated for both pressure kinds and therefore there will be present some amount of exotic matter among baryonic one.
\subsubsection{Causality condition}
To satisfy the causality condition for compact stars, sound velocity must always be lower that the speed of light (thus, $0\leq v_s\leq 1$). For the radial and tangential components, speed of sound reads:
\begin{equation}
    v_{sr}^2 = \frac{dp_r}{d\rho}=\frac{d(\alpha\rho)}{d\rho}=\alpha
\end{equation}
\begin{equation}
    v_{sr}^2 = \frac{dp_t}{d\rho}=\frac{d(\beta\rho)}{d\rho}=\beta
\end{equation}
Thus, the causality condition is always satisfied for a compact star if $0<\alpha<1$ and $0<\beta<0$.
\subsubsection{Adiabatic index}
We could check the dynamical stability of the relativistic stellar interior against infinitesimal adiabatic perturbations by following the pioneering work of Chandrasekhar \cite{Chandrasekhar1964}. Chandrasekhar predicted that for the relativistic system to be stable the adiabatic index should exceed $4/3$. This adiabatic index is defined as \cite{Maurya2017}:
\begin{equation}
    \Gamma = \frac{p_r+\rho}{p_r}\frac{dp_r}{d\rho}
    \label{eq:31}
\end{equation}
Then, for the barotropic EoS:
\begin{equation}
    \Gamma = \alpha\frac{\alpha\rho+\rho}{\alpha\rho} = 1+\alpha
\end{equation}
So the axion compact star with the barotropic EoS is dynamically stable if $\alpha>1/3$.
\subsubsection{Stability through TOV}
\begin{figure}[htbp]
    \centering
    \includegraphics[width=\textwidth]{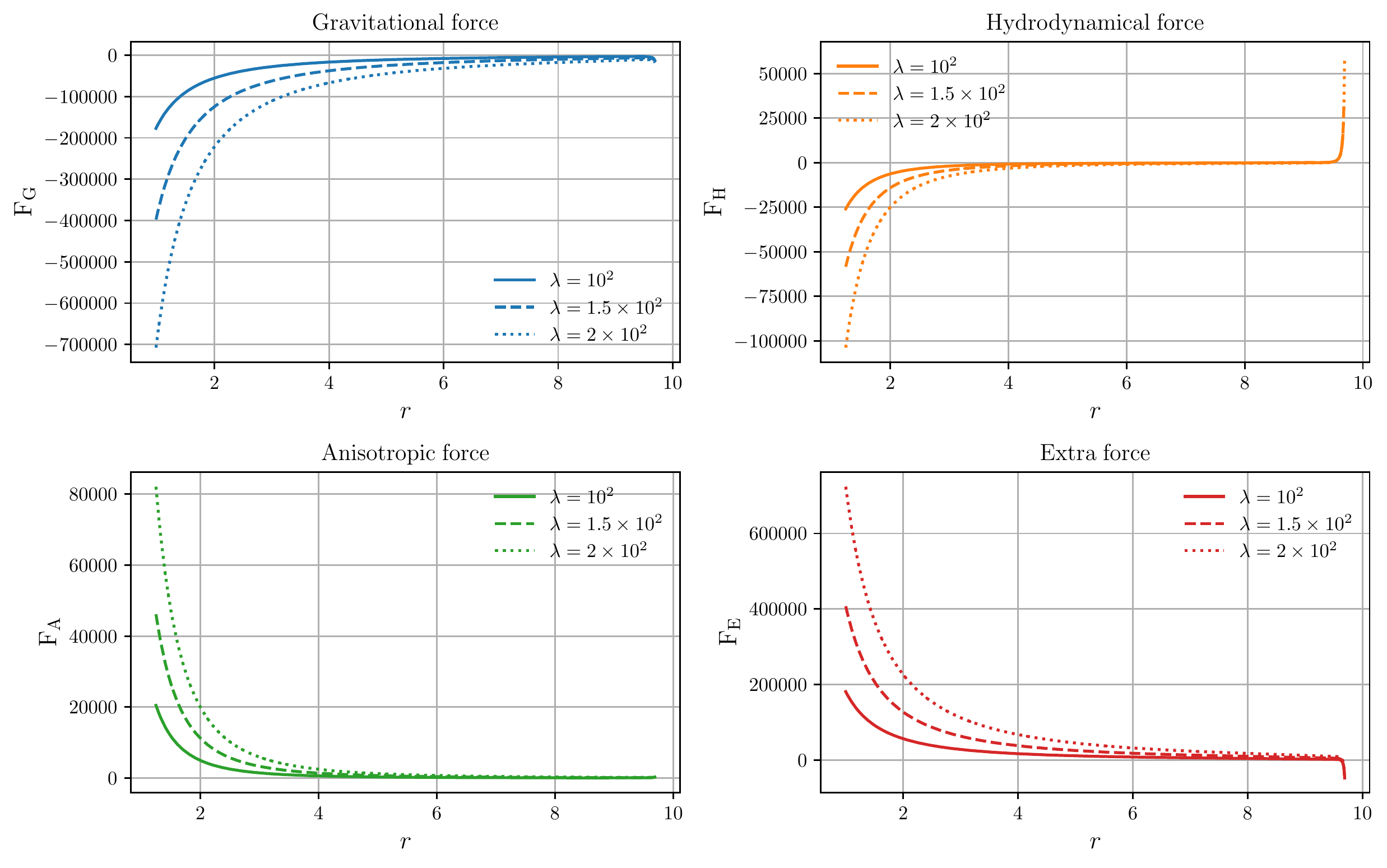}
    \caption{TOV forces for PSR J1416-223 with the barotropic EoS ($\alpha=0.5$ and $\beta=0.1$) and with varying $\lambda$, negative $B=-1$}
    \label{fig:4}
\end{figure}

We also could probe the stability of the axion compact star from the  Tolman–Oppenheimer–\newline Volkov (TOV) equation, which have the following form \cite{PhysRev.55.374,PhysRevD.78.064064}:
\begin{equation}
    -\underbrace{\frac{\Phi'}{2}(\rho+p_r)}_\text{$F_G$}-\underbrace{\frac{dp_r}{dr}}_\text{$F_H$}+\underbrace{\frac{2}{r}(p_r-p_t)}_\text{$F_A$} + F_{E}=0
    \label{eq:33}
\end{equation}
where $F_G$ is the gravitational force, $F_H$ is the hydrodynamical one and $F_A$ is the contribution to the TOV of the fluid anisotropy, finally $F_E$ is the extra force that is needed to keep the astrophysical object stable even if the stress-energy tensor isn't conserved. On the first three plots of Figure (\ref{fig:4}) we solved TOV equation (\ref{eq:33}) numerically with the varying $\lambda$ for barotropic EoS. As one could notice, axion star with barotropic EoS is stable in the region between the core and envelope.
\subsubsection{Surface redshift}
Compact star surface redshift is defined in the following way:
\begin{equation}
    \mathcal{Z}_s = |g_{tt}|^{-1/2}-1
    \label{eq:34}
\end{equation}

\begin{figure}[htbp]
    \centering
    \includegraphics[width=\textwidth]{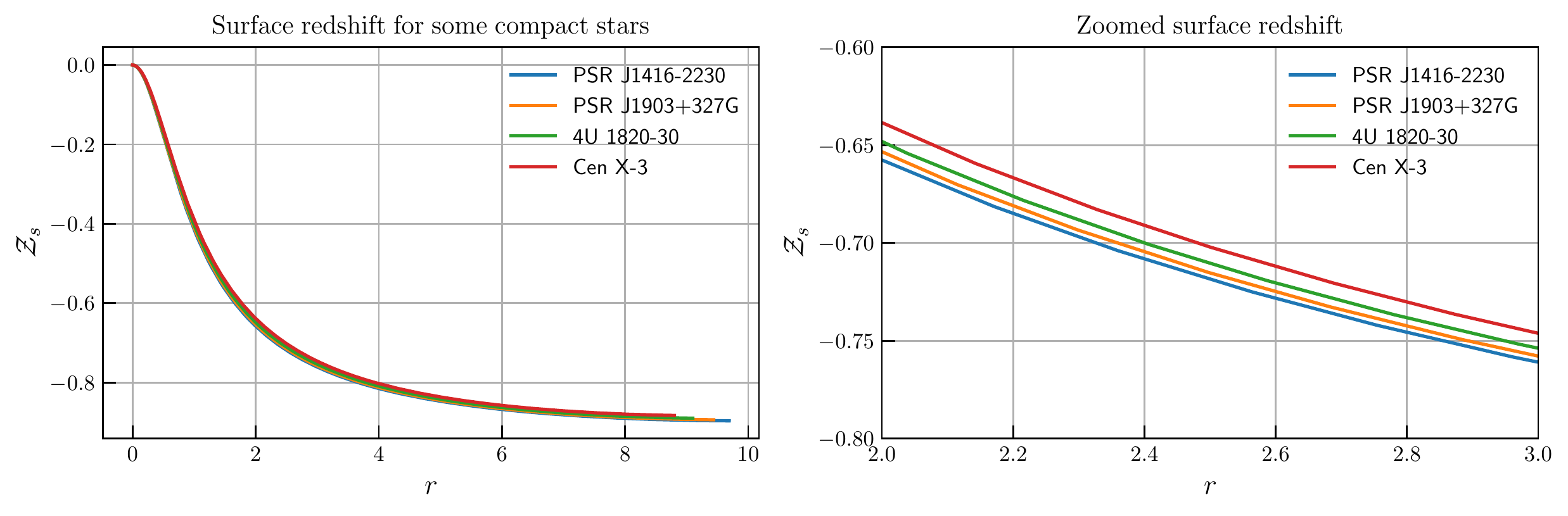}
    \caption{Surface redshift for the PSR J1416-2230, PSR J1903+327, 4U 1820-30 and Cen X-3 in the model with the barotropic EoS ($\alpha=0.5$, $\beta=0.1$) and for negative $B=-1$}
    \label{fig:444}
\end{figure}

Values of the surface redshift must not exceed 2. On the plot of Figure (\ref{fig:444}) we solved equation (\ref{eq:34}) for 4 compact stellar objects PSR J1416-223, PSR J1903+327, 4U 1820-30 and Cen X-3 (for the values of mass and radius refer to the Table (\ref{tab:1})). It is obvious that for this stars surface redshift is lower than $2$, which is one of the necessary conditions, that must be obeyed.
\subsection{Isotropic Quark-axion star with MIT bag EoS}
\subsubsection{Energy conditions}
One could assume that the asymptotically free quarks are trapped in a region with finite volume, which is called bag. The bag constant $\mathcal{B}$ exist as a inward pressure that traps the quarks inside the bag. In our particular case, we assume the so-called MIT bag EoS, in which for simplicity it is assumed that the up, down and strange quarks are non-interacting ones and massless. Thus, in the MIT bag model, radial pressure is given by:
\begin{equation}
    p_r = \sum_{f=u,d,s}p^f - \mathcal{B}
\end{equation}
and energy density consequently is:
\begin{equation}
    \rho = \sum_{f=u,d,s}\rho^f  + \mathcal{B}
\end{equation}
where the $p^f$ is the radial pressure for each flavor, $\rho^f$ is the energy density for each flavor. By combining the equations above, we could get the well-known simplified MIT bag EoS model:
\begin{equation}
    p_r = \frac{1}{3}(\rho - 4\mathcal{B})
    \label{eq:37}
\end{equation}
\begin{figure}[!htbp]
    \centering
    \includegraphics[width=\textwidth]{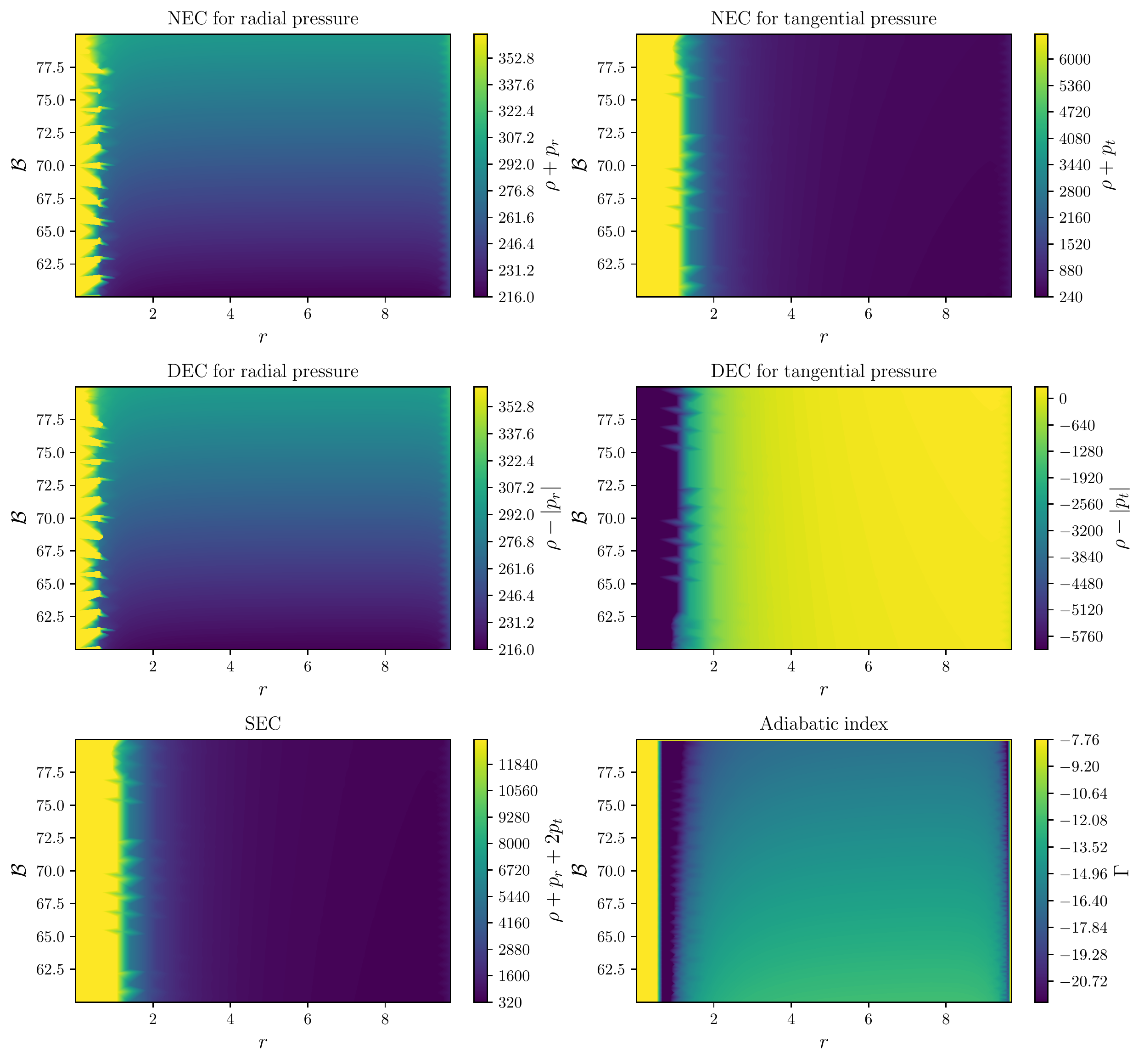}
    \caption{(first three plots) Isotropic Null, Dominant and Strong energy conditions for the PSR J1416-2230 in the presence of two-fields axion with the MIT bag EoS, (last plot) Adiabatic index for the Quark-axion compact star with the MIT bag equation of State}
    \label{fig:2}
\end{figure}
\newline We plotted energy conditions for the MIT bag EoS model with anisotropic pressures on the first five plots of the Figure (\ref{fig:2}). Same as for the barotropic EoS, null and strong energy conditions are satisfied (we vary the $\mathcal{B}$ in the limits $60\leq\mathcal{B}\leq80$), but dominant energy condition is violated (could be satisfied for small $|\lambda|$, but in that case NEC will be disobeyed). 
\subsubsection{Causality condition}
Isotropic causality condition reads:
\begin{equation}
    0<\frac{dp}{d\rho}<1\Rightarrow \frac{dp}{d\rho}=1/3
\end{equation}
Therefore causality condition is satisfied for every value of MIT bag constant $\mathcal{B}$.
\subsubsection{Adiabatic index}
Using the equation (\ref{eq:31}) for the isotropic fluid one could obtaine following adiabatic index with MIT bag EoS:
\begin{equation}
    \Gamma = \frac{4}{3}+\frac{4\mathcal{B}}{-4\mathcal{B}+\rho}\Rightarrow\frac{4\mathcal{B}}{-4\mathcal{B}+\rho}>0
    \label{eq:38}
\end{equation}
\newline We depicted the numerical solution for the equation (\ref{eq:38}) on the last plot of the Figure (\ref{fig:2}). As we discovered during the numerical analysis, star is unstable from the adiabatic perturbations everywhere, even at the origin.
\subsubsection{Probing the Quark-axion star stability from the TOV}
\begin{figure}[!htbp]
    \centering
    \includegraphics[width=\textwidth]{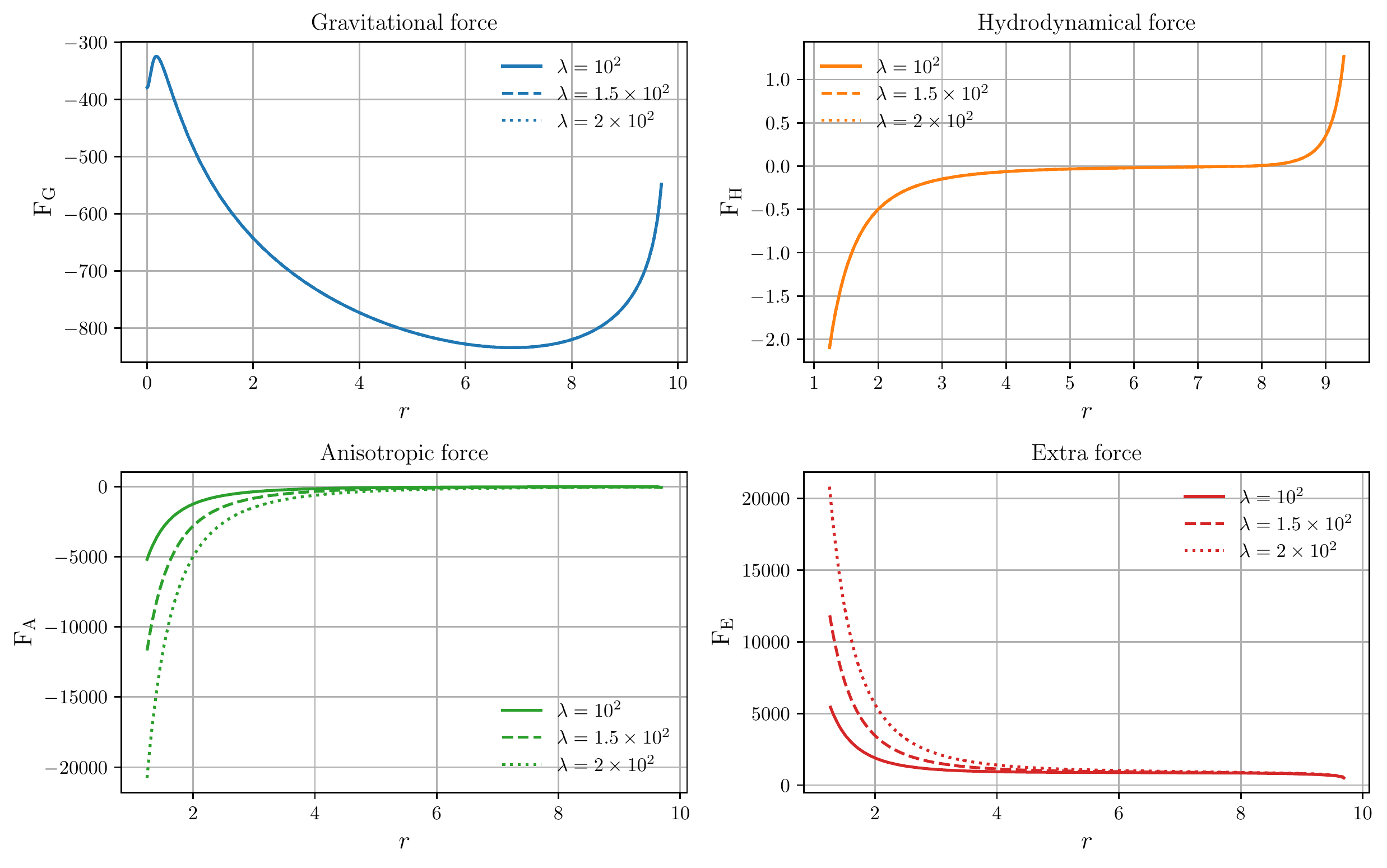}
    \caption{TOV forces for PSR J1416-223 with the MIT bag EoS ($\mathcal{B}=60$) and with varying $\lambda$, negative $B=-1$}
    \label{fig:7}
\end{figure}

We have placed the numerical solutions for the TOV equation above with the varying values of parameter $\lambda$ on the Figure (\ref{fig:7}). It is worth mentioning that the TOV forces grow as $|\lambda|\to\infty$.

\section{Axion stars with the complex scalar field} \label{sec:3}
For the theory with the two-fields axion and coupled to gravity complex scalar field we have the following EH action integral:
\begin{equation}
    \mathcal{S}[g,\Gamma,\psi_I,\phi,\Psi_i] = \int_\mathcal{M}d^Dx \sqrt{-g}\bigg[\kappa(\mathcal{R}-2\Lambda)-\nabla_\mu\overline{\Phi}\nabla^\mu\Phi - V(\Phi\overline{\Phi}) -\frac{1}{2}\sum^2_{I=1}(\partial^\mu\psi_I\partial_\mu\psi_I)+\mathcal{L}(\Psi_i)\bigg]
\end{equation}
where $\Phi$ is the complex coupled to the gravity scalar field ($\overline{\Phi}$ is the complex conjugate of the scalar field). We also could define the potential (with the quartic self-interaction) of the given scalar field:
\begin{equation}
    V(\overline{\Phi}\Phi) = \frac{1}{2}\mu^2\overline{\Phi}\Phi+\frac{1}{4}\chi(\overline{\Phi}\Phi)^2
\end{equation}
where $\mu=mc/\hslash$ is the mass of bosonic particle (from the beginning we have assumed that $c=\hslash=1$) and $\chi$ is the self-interaction coupling constant. Then, by varying the action w.r.t. metric tensor and $\Phi$ we could obtain:
\begin{equation}
    \kappa(G_{\mu\nu}+\Lambda g_{\mu\nu})=\frac{1}{2}(T^\Phi_{\mu\nu}+T^\psi_{\mu\nu}+T_{\mu\nu}^{\mathrm{M}})
\end{equation}
\begin{equation}
    \nabla_\mu\nabla^\mu\Phi =\frac{\partial V}{\partial \Phi}= \mu^2 \Phi + \chi |\Phi|^2\Phi
    \label{eq:44}
\end{equation}
Consequently, stress-energy tensor of $\Phi$ is:
\begin{equation}
    T_{\mu\nu}^\Phi = -\frac{1}{2}g_{\mu\nu}\partial_\alpha \overline{\Phi}\partial_\alpha\Phi - V(\Phi) + \frac{1}{2}(\partial_\mu\overline{\Phi}\partial_\nu\Phi + \partial_\mu\Phi\partial_\nu\overline{\Phi})
\end{equation}
\subsection{Complex scalar field}
Firstly, to solve the field equations we might specify the complex scalar field. We will numerically derive the scalar field directly from the Klein-Gordon equation (\ref{eq:44}) using assumption of complex scalar field periodic evolution \cite{Boyadjiev2001MathematicalMO}:
\begin{equation}
    \Phi(t,r)=\phi(r)e^{i\omega t}
\end{equation}
Where $\phi(r)$ is real scalar function which depends only on the radial coordinate $r$. We are going to derive $\Phi(t,r)$ numerically using initial conditions $\phi(R)=10^{-1}$ and $\phi'(R)=0$ and Runge-Kutta ODE solver of 4th order. As we noticed, to obtain physically viable solutions for $\phi$ defined on the whole domain $r$ we need to impose that there is no self interaction, $\chi=0$.
\subsection{EFE's for the axion stars with the complex scalar field}
Einstein field equations for the axion stars with the complex scalar field coupled to gravity are calculated below:
\begin{equation}
    \begin{gathered}
    \rho =-\frac{1}{4} \Bigg(\frac{\phi (r)^2 \left(-2 \omega ^2 \left(A r^2+B r^4+3\right)+2 \mu
   ^2+\chi  \phi (r)^2\right)}{A r^2+B r^4+1}+\frac{1}{r^2}\bigg(8 \bigg(\frac{e^{-C r^2}
   \left(2 C r^2-1\right)}{D^2}-\Lambda  r^2+1\bigg)\bigg)-2 \phi
   '(r)^2\Bigg)
    \end{gathered}
\end{equation}
\begin{equation}
    \begin{gathered}
    p_r =-\frac{1}{4 D^2}\Bigg(e^{-C r^2} \Bigg(\frac{1}{r^2}\bigg(8 \left(\frac{2 A r^2+4}{A r^2+B r^4+1}+D^2
   \left(-e^{C r^2}\right) \left(\Lambda  r^2-1\right)-5\right)\bigg)\\
   -2 \phi (r)^2
   \left(D^2 \omega ^2 e^{C r^2}+\phi ^2\right)+\left(4-2 D^2 e^{C
   r^2}\right) \phi '(r)^2-\chi  \phi (r)^4\Bigg)\Bigg)
    \end{gathered}
\end{equation}
\begin{equation}
    \begin{gathered}
    p_t = -\frac{1}{4} \Bigg(\bigg(8 e^{-C r^2} \left(-2 r^4 \left(A^2 C-3 A B+2 B
   C\right)+r^2 \left(A^2-3 A C+8 B\right)+B r^6 (4 B-5 A C)+2 A-3 B^2
   C r^8-C\right)\bigg)\\
   \bigg/\bigg(D^2 \left(A r^2+B r^4+1\right)^2\bigg)+8 \Lambda -\frac{2
   \lambda ^2}{r^2}+2 \phi (r)^2 \left(\frac{\mu ^2}{r^2}+\omega ^2\right)+2 \phi
   '(r)^2+\frac{\chi  \phi (r)^4}{r^2}\Bigg)
    \end{gathered}
\end{equation}
Therefore, we could proceed and renew junction condition now with the present complex scalar field.
\subsection{Junction conditions}
Because of the presence of complex scalar field, expressions $p_r$ changed, and so $C$ and $D$ variables changed ($A$, $B$ remain the same). Therefore, we need to derive the parameter $D$  from the condition $p_r(R)=0$ (general view of $C$ does not change and looks exactly like (\ref{eq:25})):
\begin{equation}
\begin{gathered}
    D=\bigg(\left(6 M+\Lambda  R^3-3\right) \bigg(24 \left(2 B R^5+6 M-\Lambda  R^3+4 R-3\right)+R^2 \left(6 M-\Lambda  R^3+6 R-3\right)\\
    \times\left(2 \mu^2 \phi(R)^2-4 \phi'(R)^2+\chi  \phi(R)^4\right)\bigg)\bigg)\bigg/\bigg(6 R \left(-6 M+\Lambda  R^3-6
   R+3\right) \left(4 \Lambda  R^2+R^2 \left(\phi'(R)^2+\omega ^2 \phi(R)^2\right)+4\right)\bigg)
   \end{gathered}
\end{equation}
Numerical solutions for T-K spacetime coefficients are located on the Table (\ref{tab:2}). Then, while we already defined all of the necessary quantities, we could go further to the barotropic EoS investigation.
\begin{table}[]
\centering
\resizebox{0.8\textwidth}{!}{%
\begin{tabular}{llllllll}
\hline
Star            & $M (M_\odot)$ & $R$(km) & $A$         & $B$            & $C$          & $D$        & Reference                                                \\ \hline
PSR J1416-2230  & 1.97          & 9.69    & $1.95284$  & $-0.01$ & $-0.0590632$  & $0.44318$ & \cite{2010Natur.467.1081D}              \\
PSR J1903+327   & 1.667         & 9.438   & $1.90476$ & $-0.01$ & $-0.0412337$ & $0.400204$ & \cite{2011MNRAS.412.2763F}              \\
4U 1820-30      & 1.58          & 9.1     & $1.84304$ & $-0.01$ & $-0.04224$ & $0.344252$ & \cite{2010}                             \\
Cen X-3         & 1.49          & 8.50    & $1.73956$ & $-0.01$ & $-0.0444363$ & $0.251992$ & \cite{2011ApJ...730...25R}              \\
EXO 1785-248    & 1.3           & 8.99    & $1.82278$ & $-0.01$ & $-0.0372638$ & $0.32734$ & \cite{Ozel2008}                         \\
SAX J1808.43658 & 1.32          & 4.14    & $1.25285$ & $-0.01$      & $-0.0992585$  & $0.128054$ & \cite{10.1111/j.1365-2966.2009.14562.x} \\ \hline
\end{tabular}%
}
\caption{T-K spacetime coefficients $A$, $B$, $C$ and $D$ for some compact stars (we assumed that the AdS radius is $l=1$ and thus $\Lambda=-3$, also we consider the case with the negative $B=-1$) coupled to the complex scalar field (with $\mu^2=6.25$ and $\omega=10$)}
\label{tab:2}
\end{table}
\subsection{Barotropic axion stellar objects}
In this subsection we will investingate the axion stars with complex scalar field coupled to gravity and with the barotropic equation of state.
\subsubsection{Energy conditions}
We firstly as usual want to probe the energy condition of our Boson-axion star. As we already stated, we will use the barotropic EoS from equations (\ref{eq:26}) and (\ref{eq:27}).
\begin{figure}[!htbp]
    \centering
    \includegraphics[width=\textwidth]{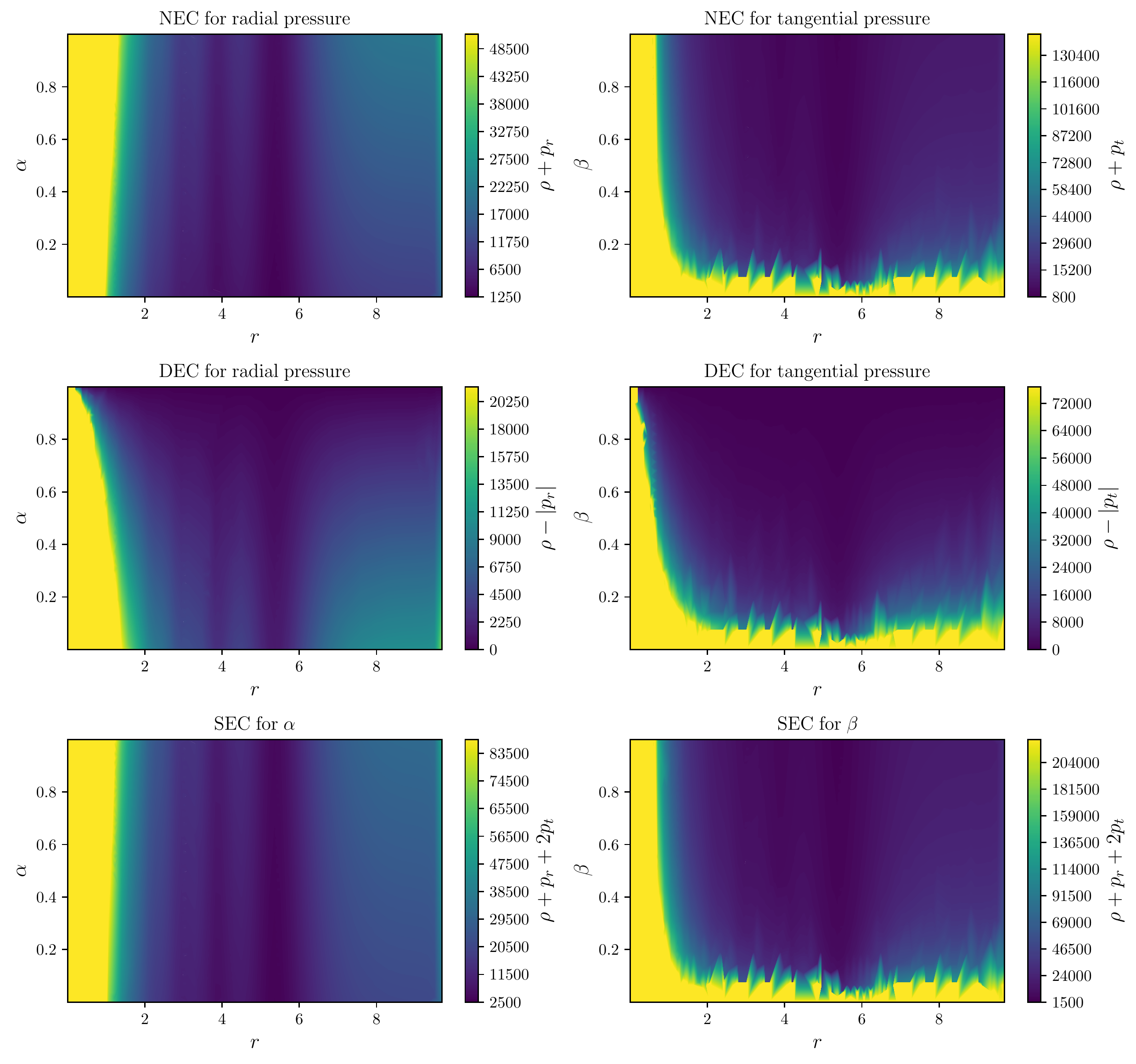}
    \caption{Null, Dominant and Strong energy conditions for the PSR J1416-2230 in the presence of two axion and one complex scalar fields. For simplicity we also assume that, $\omega=10$, $\lambda=10^2$ and $B=-1$. At final, complex scalar field mass is $\mu^2=6.25$}
    \label{fig:8}
\end{figure}

We have placed the numerical solutions for various energy conditions at the Figure (\ref{fig:8}). Judging by the data, obtained from the numerical analysis we came to the conclusion that NEC, DEC and SEC are generally satisfied. Also, it is worth to notice that in relation to the case without complex scalar field, EC's have slightly bigger values.
\subsubsection{TOV stability}
\begin{figure}[!htbp]
    \centering
    \includegraphics[width=\textwidth]{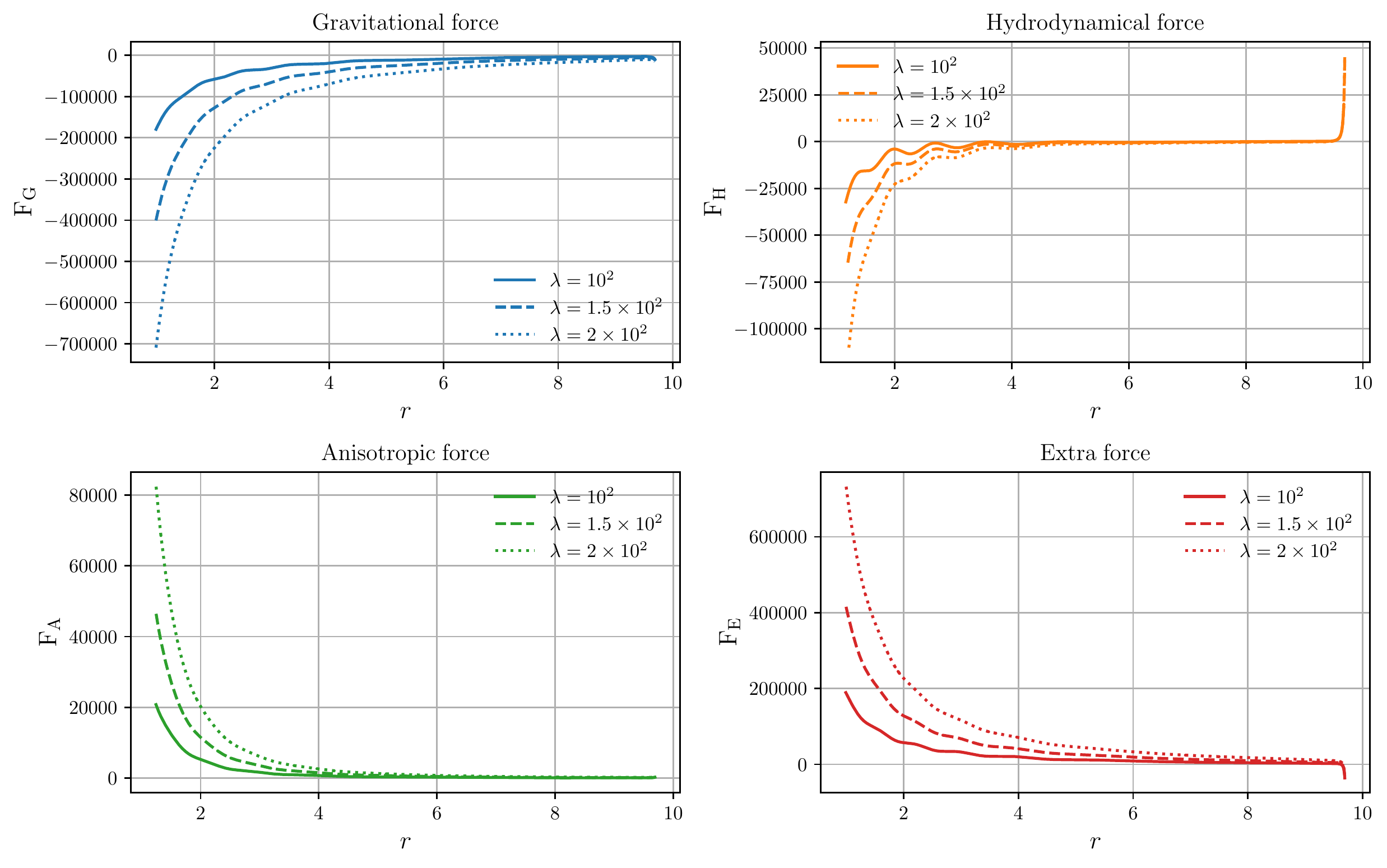}
    \caption{TOV forces for PSR J1416-223 with the barotropic EoS ($\alpha=0.5$ and $\beta=0.1$) and with varying $\lambda$, coupled to gravity complex scalar field (as usual for simplicity we set $\omega=10$ and $B=-1$)}
    \label{fig:9}
\end{figure}

We show the TOV forces for the barotropic axion star in the vicinity of complex scalar field on the four plots of Figure (\ref{fig:9}). As we see, TOV forces are very similar to the the barotropic axion star case without the complex scalar field, but in addition they have oscillating behavior. But if we will assume bigger values of scalar field mass, $|\mathrm{F_{G,H,A,E}}|\to\infty$ (general view of forces profile will be conserved in that case)

\subsection{Axion stars with the MIT bag EoS}
\subsubsection{Energy conditions}

\begin{figure}[!htbp]
    \centering
    \includegraphics[width=\textwidth]{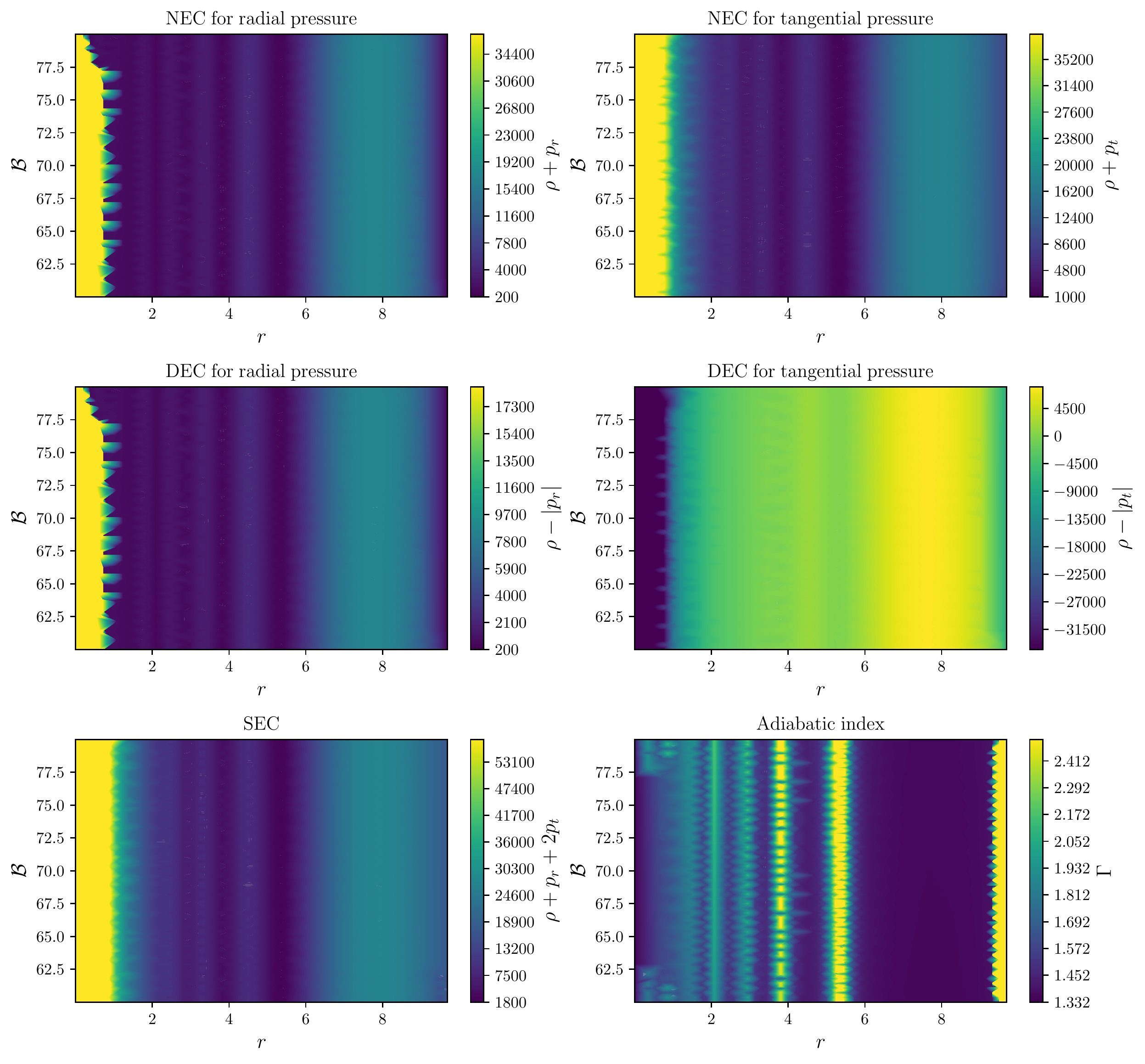}
    \caption{(first five plots) Null, Dominant and Strong energy conditions for the axion star with the MIT bag EoS and with $\omega=10$ and $B=-1$, $\mu^2=6.25$, (last plot) Adiabatic index for the Boson-Quark-axion compact star with the MIT bag Equation of State}
    \label{fig:10}
\end{figure}

Therefore, energy conditions for the anisotropic pressure are placed on the first three plots of Figure (\ref{fig:10}).  As we see, all of the energy conditions except the DEC for tangential pressure are validated.
\subsubsection{Adiabatic index}
General view of the adiabatic index for the MIT bag EoS doesn't change and thus we could derive adiabatic index directly from the equation (\ref{eq:38}). 
Consequently, we plot adiabatic index for the Boson-Quark-Axion (further - BQA) compact star at the last plot of Figure (\ref{fig:10}). As you may obviously notice from the figure data, $\Gamma>4/3$ everywhere up to $r\approx 5$ and consequently, our solution is dynamically stable within the stellar core. Also, similar to the energy conditions, $\Gamma>4/3$ holds even for every $\mu^2\geq0$.
\subsubsection{TOV stability}
From Figure (\ref{fig:77}) we could state that as well as for the adiabatic stability, from the modified TOV equation we could say that without the presence of additional TOV force matter located in the axion star interior is unstable. As well, if $\mu^2\to\infty$, then $|\mathrm{F_{G,H,A,E}}|\to\infty$.
\section{Conclusions} \label{sec:4}
\begin{figure}[!htbp]
    \centering
    \includegraphics[width=\textwidth]{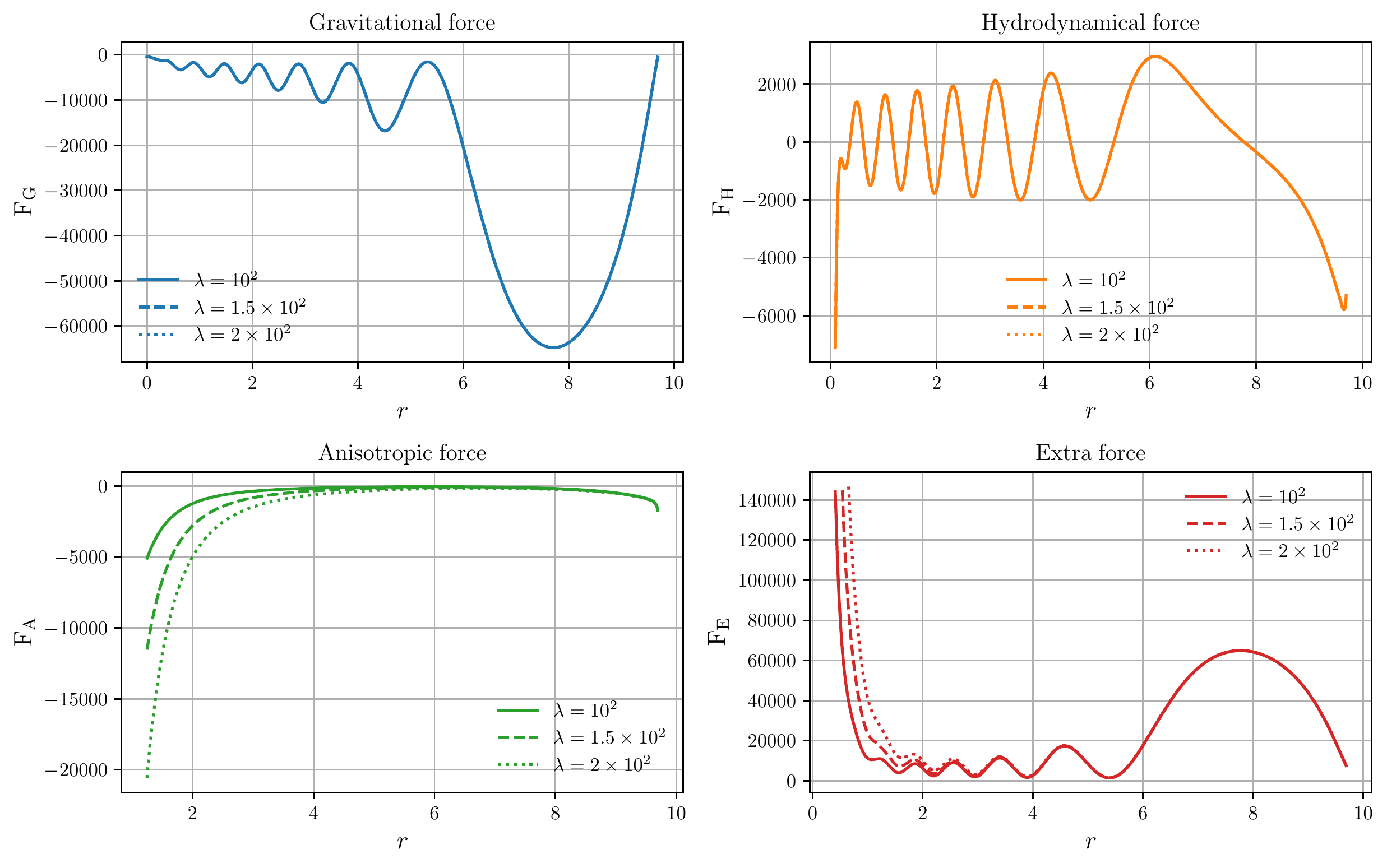}
    \caption{TOV forces for PSR J1416-223 with the MIT bag EoS ($\mathcal{B}=60$) and with varying $\lambda$, coupled to gravity complex scalar field (with $\mu^2=6.25$ and $\omega=10$), negative $B=-1$}
    \label{fig:77}
\end{figure}

In the present article we have studied the static and spherically symmetric non-charged compact stars in the presence of Dante's inferno axion monodromy model with/without the complex scalar field, with the negative $\Lambda$ term. In this section we want to summarize all of the key results, that was obtained in the
paper.
For the axion stars without the complex scalar field:
\begin{itemize}
    \item \textbf{Energy conditions}: for the barotropic EoS, Null, Dominant and Strong energy conditions were satisfied at every point of the compact star interior spacetime and for MIT bag EoS NEC and SEC were satisfied and DEC was violated for tangential pressure (both cases admit that $|\lambda|\gg0$, for the further details, check Figures (\ref{fig:1}) and (\ref{fig:2})).
    \item \textbf{Causal condition}: for the barotropic axion star causal condition was satisfied with $\alpha\land\beta\in(0,1)$, for the MIT bag causal condition was satisfied for every value of $\mathcal{B}$
    \item \textbf{Adiabatic index}: for the barotropic EoS, adiabatic index was $\Gamma=1+\alpha$, and thus for the axion star to be stable of adiabatic perturbations, values of $\alpha$ must exceed $1/3$. On the other hand, for the MIT bag EoS, matter was unstable everywhere in the axion star interior, even at the stellar core (see the last plot of Figure (\ref{fig:2})).
    \item \textbf{Surface redshift}: surface redshift function is depicted for some compact stars at the Figure (\ref{fig:444}). $\mathcal{Z}_s$ do not exceed $2$ as needed.
    \item \textbf{TOV stability}: For the barotropic EoS, forces present in the TOV are depicted on the Figure (\ref{fig:4}), for the MIT bag EoS on the Figure (\ref{fig:7}).
\end{itemize}
For the axion stars with the coupled to gravity complex scalar field:
\begin{itemize}
    \item \textbf{Energy conditions}: if we consider complex scalar field as an additional field coupled to gravity, then for the axion stars with the barotropic equation of state energy conditions are satisfied everywhere if EoS parameters are in the permitted region $(0,1)$ and if $|\lambda|\gg0$. In turn, for the MIT bag EoS every EC except tangential DEC was validated everywhere (energy conditions for the MIT bag EoS are plotted on the Figure (\ref{fig:10})).
    \item \textbf{Causal condition}: causal conditions are the same as for the case without the complex scalar field
    \item \textbf{Adiabatic index}: adiabatic index for the barotropic EoS remain the same, but for the MIT bag in relation the the case without the complex scalar field, matter is unstable near the envelope.
    \item \textbf{Surface redshift}: junction conditions for $g_{tt}$ component remain the same as for the case without complex scalar field, so surface redshift still does not exceed 2.
    \item \textbf{TOV stability}: for the barotropic EoS, TOV forces are presented on the Figure (\ref{fig:9}), for the MIT bag one on the Figure (\ref{fig:77}).
\end{itemize}
\section*{Acknowledgment}
We are thankful to the honorable anonymous referee and the editor for helpful comments, which have significantly improved our work in terms of research quality and presentation. 

\end{document}